# Presenting the Multi-Objective Optimization Model of Search and Rescue Network

Md Mashum Billal[1] [0000-0001-5557-3613] and Maryam Maleki[2] [0000-0002-9895-7999]

[1] Department of Mechanical Engineering (Engineering Management), University of Alberta, Edmonton, Canada
mdmashum@ualberta.ca
[2] Systems Engineering Department, University of Arkansas at Little Rock, Little Rock, AR, 72204, USA
mmaleki@ualr.edu

**Abstract.** The Search and Rescue Network (SAR) is a kind of emergency network that pursuit people in need or imminent danger. This paper aims using a priori optimization to demonstrate the optimal assignment of HFDF receivers to the Generalized Search and Rescue (GSAR) network, which is independent of the weighting of the transmitter areas. The mathematical model seeks two objectives, the first one is maximizing the expected number of LOBs for HFDF receivers. The second is providing a fair share number of HFDF receivers allowed to cover the frequency. The result shown the efficiency of presented model ran by CPLEX toolbox of MATLAB 2020 software.

## 1 Introduction

Search and rescue network has different forms and each of them with unique risks and dangers to victim and responder (Murphy et al. 2008; Razi and Karatas 2016).The U.S. is founded and maintained a system of search and rescue (SAR) stations encompassing seas and oceans, these stations are responsible for receiving and processing signals from distressed ships, vessels and airplanes in order to initiate the emergency operations. The spark of any emergencies is the time when three or more stations receive and process the same distress signal since in order to find an approximation of distressed vessel three stations are required.

### 1.1 Relation Between Receiving Subsystems (RS) and Central Control (CC)

It is noted that each station in the SAR network has only one RS system, but the number of high frequency direction finding (HFDF) receivers in each station is different. In our problem, the number of HFDF receivers varies between 0 and 10. For more clarification, RS probes the entire frequency spectrum and has less sensitive and accurate than the HFDF. Moreover, RS has the limitation on small signal-to-noise ration unlike HFDF. Every HFDF receivers is allotted to a 1 MHz bund within the frequency spectrum.

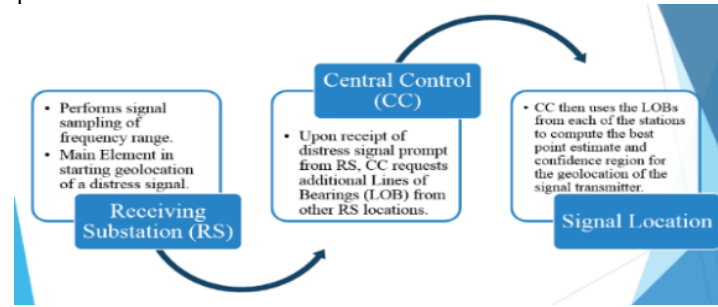

**Figure 1. Relationship between RS and CC**

Prior to research background, the following schematic figure depicts the area of research in accordance with the receiving substation and estimated point of transmitted point and an acceptable circularized error radius.

## 1.2 Error radius

The most recent research history on the topic of search and rescue culminated in optimal methods for the location of stations and frequency assignments. Since the subject is a well-acknowledged area of research, many previous researches had added to the body of knowledge. (Jr 1992) thoroughly discussed the problem and different analytical approaches. He explained that "classical sensitivity analysis and tolerance analysis were used to analyze the frequency assignments generated by the different weight sequences. The weight sequence with all weights having equal value produced the most robust frequency assignments for all time blocks".
We followed the same footpath to recalculate the results once more time. Although (Jr 1992) used ADBASE, LINGO and CPLEX IBM ILOG Studio also provided the same computer runs results, which are enclosed to this report as well.

As by (Jr 1992) cited, the basis of his research is founded mainly on two antecedences, first, Steppe used a two-stage, network-flow multi-Objective linear integer programming (MOLIP) model to determine the optimal position of the stations for the SAR problem. Second, Johnson's further work in this field has established optimal frequency assignments using the MOLIP network-flow model.

## 1.3 Research Objectives

The goal of this research is to use a priori optimization to show that the optimal assignment of HFDF receivers to the Generalized Search and Rescue (GSAR) network is independent of the weighting of the transmitter areas. This is achieved by examining the impact of changing the weight value of a specific transmitter area on the geolocation likelihood for that area. The mathematical model has two purposes, the first of which is to optimize the predicted number of LOBs for HFDF receivers. And the second is to have a reasonable share of the number of HFDF receivers allowed to cover the frequency (Hayat et al. 2017).

## 2 Model Formulation

This section is an overview of the multi-objective linear programming (MOLP) and network programming formulas for the search and rescue network (Arani et al. 2020; MomeniTabar et al. 2020), (Razi and Karatas 2016). The weights for transmitter areas shall be given by the Department of Defense (DOD).

The notations, parameters, and variables are:

$i$: transmitter locations

$j$: receiving locations

$k$: frequency bands

$F_{ik}$: Probability of a distress signal from location i on frequency k

$P_{ijk}$: Probability that a distress signal from location i on frequency k is acquired by station j

$W_{ij}$: Probability that a line of bearing from station j is within the acceptable circularized error region defined for location i

$U_i$: The normalized weight (0 – 1 range) of a distress signal from location i

TN: The total number of HFDF receivers

FS: The fair share of HFDF receivers for each frequency.

Where FS is the integer greater than or equal to the total number of HFDF receivers divided by the total number of frequencies to be covered.

$$X_{jk} = \begin{cases} 1 \\ 0 \end{cases}, \text{if station j is assigned cover frequency k, otherwise 0}$$

$$Y_k = \begin{cases} n \\ 0 \end{cases}, \text{if frequency k has excess coverage by n stations, otherwise 0}$$

### 2.1 Objectives and Constraints

The model formulation for this multi-objective optimization model of a search and rescue network takes the following form:

Objective Function 1: This objective function maximizes the estimated number of accurate bearing lines for HFDF receivers (Razi and Karatas 2016).

$$Max \sum_i \sum_j \sum_k U_i W_{ij} F_{ik} P_{ijk} X_{jk}$$

Objective Function 2: This objective function minimizes the excess coverage of HFDF receivers for each frequency.

$$Min \sum_k Y_k$$

Constraint 1: Limit the number of HFDF frequency assignments at each station to the number of receivers located at each station.

$$\sum_k X_{jk} \le m_j, \qquad \forall j$$

Constraint 2: This restriction allows at least two HFDF receivers to be allocated to cover each frequency.

$$\sum_j X_{jk} \ge 2, \qquad \forall k$$

Constraint 3: Determines the sum of excess coverage provided at each frequency. The vector $Y_k$ is the indicator of excess coverage.

$$\sum_j X_{jk} - Y_k \le FS, \qquad \forall k$$

## 2.2  Obtained Data from DOD

A case study of actual data is provided, and the results are regenerated since the software changed. The following data is used to calculate the weights for the problem. Table 1 provides the probability of a signal being transmitted by a transmitter i on frequency k.

**Table 1.** Signal Transmission and frequency probability.

| i / k | Frequency 1 | Frequency 2 | Frequency 3 |
|---|---|---|---|
| Transmitter 1 | 0.04 | 0.04 | 0.04 |
| Transmitter 2 | 0.00 | 0.00 | 0.01 |
| Transmitter 3 | 0.03 | 0.05 | 0.05 |
| Transmitter 4 | 0.00 | 0.00 | 0.00 |

Table 2 indicates the likelihood of a signal being transmitted from transmitter i to frequency k and acquired by station j.

**Table 2.** probability of signal transmission and station acquisition.

| j | Transmitter | | | | Transmitter | | | | Transmitter | | | |
|---|---|---|---|---|---|---|---|---|---|---|---|---|
| | 1 | 2 | 3 | 4 | 1 | 2 | 3 | 4 | 1 | 2 | 3 | 4 |
| 1 | 0. | 0.32 | 0.51 | 0.01 | 0.95 | 0.13 | 0.35 | 0.0 | 0.96 | 0.33 | 0.52 | 0.01 |

| 2 | 0. | 0.44 | 0.13 | 0.01 | 0.98 | 0.08 | 0.01 | 0.0 | 0.98 | 0.30 | 0.01 | 0.01 |
| 3 | 0. | 0.01 | 0.01 | 0.01 | 0.92 | 0.46 | 0.71 | 0.0 | 0.83 | 0.31 | 0.51 | 0.01 |
| 4 | 0. | 0.97 | 0.01 | 0.01 | 0.98 | 0.01 | 0.12 | 0.0 | 0.90 | 0.01 | 0.01 | 0.01 |
| 5 | 0. | 0.02 | 0.01 | 0.01 | 0.94 | 0.04 | 0.01 | 0.0 | 0.94 | 0.19 | 0.0 | 0.01 |

Table 3 indicates the likelihood that station j will receive a signal from the transmitter I when a signal has been transmitted.

**Table 3.** Probability of station receipt of signal

| i / i | Station 1 | Station 2 | Station 3 | Station 4 | Station 5 |
|---|---|---|---|---|---|
| Transmitter 1 | 0.3808 | 0.747 | 0.1951 | 0.121 | 0.7956 |
| Transmitter 2 | 0.1477 | 0.1301 | 0.1140 | 0.0596 | 0.2504 |
| Transmitter 3 | 0.1471 | 0.0892 | 0.1580 | 0.0834 | 0.1509 |
| Transmitter 4 | 0.0515 | 0.7679 | 0.0615 | 0.0820 | 0.0427 |

Table 4 offers different weighting sequences for the nine solutions to the sample problem.

**Table 4.** Weighting sequence for the nine solutions to the sample problem.

| i / | Stat | Statio | Statio | Statio | Statio | Statio | Statio | Statio | Statio |
|---|---|---|---|---|---|---|---|---|---|
| 1 | 0.25 | 0.50 | 0.167 | 0.167 | 0.167 | 0.70 | 0.10 | 0.10 | 0.10 |
| 2 | 0.25 | 0.167 | 0.50 | 0.167 | 0.167 | 0.10 | 0.70 | 0.10 | 0.10 |
| 3 | 0.25 | 0.167 | 0.167 | 0.50 | 0.167 | 0.10 | 0.10 | 0.70 | 0.10 |
| 4 | 0.25 | 0.167 | 0.167 | 0.167 | 0.50 | 0.10 | 0.10 | 0.10 | 0.70 |

**Table 5.** Manual sensitivity analysis range for time block one weight

| Weight # | Original Value | Low Value | High Value |
|---|---|---|---|
| 20 | 0.203 | 1% | 10% |
| 22 | 0.145 | 16% | 4% |
| 27 | 0.203 | 6% | 0% |
| 30 | 0.145 | 28% | 5% |
| 31 | 0.203 | 11% | 8% |

**Table 6.** Manual sensitivity analysis range for time block six weights

| Weight # | Original Value | Low Value | High Value |
|----------|---------------|-----------|------------|
| 9 | 0.1491 | 10% | 15% |
| 20 | 0.1491 | 23% | 4% |
| 27 | 0.1491 | 3% | 27% |
| 30 | 0.1355 | 28% | 4% |
| 31 | 0.1897 | 11% | 14% |
| 40 | 0.1355 | 2% | 13% |

### 3.2 Methodology

The technique used was a constraint reduced feasible region method in a "toy problem" type of scenario where a condensed version of the larger problem was extracted and run to show that the calculations are accurate, and that the solution is viable.

The constraint reduced method to solve a MCLP is to "convert one of the two criterion functions, in this case f2(x), into a constraint, which is added to the existing constraint set $x \in X$." (Arani et al. 2020; Mehrez et al. 1996), (Hayat et al. 2020), (Nielsen et al. 2021). The formulation of our toy problem therefore goes from the following objective function and constraint function notation:

$$max\, f_1(x) = 0.0043 * x11 + 0.004275 * x12 + 0.004725 * x13$$
$$+ 0.007325 * x21 + 0.00726 * x22 + 0.00736 * x23$$
$$+ 0.001938 * x31 + 0.0032 * x32 + 0.002725 * x33$$
$$+ 0.001183 * x41 + 0.0013 * x42 + 0.001103 * x43$$
$$+ 0.007813 * x51 + 0.007495 * x52 + 0.0076 * x53;$$
$$min\, f_2(x) = -y1 - y2 - y3;$$
$$x11 + x12 + x13 + x21 + x22 + x23 + x31 + x32 + x33 + x41 + x42 + x43 + x51$$
$$+ x52 + x53 <= 15;$$
$$X11 + x21 + x31 + x41 + x51 - e1 <= 3;$$
$$x12 + x22 + x32 + x42 + x52 - e2 <= 3;$$
$$X13 + x23 + X33 + X43 + x53 - e3 <= 3;$$
$$X11 + x21 + x31 + x41 + x51 >= 2;$$
$$X12 + X22 + X32 + X42 + X52 >= 2;$$
$$X13 + x23 + x33 + x43 + x53 >= 2;$$

To the following form:

$$max\, f_1(x) = 0.0043 * x11 + 0.004275 * x12 + 0.004725 * x13$$
$$+ 0.007325 * x21 + 0.00726 * x22 + 0.00736 * x23$$
$$+ 0.001938 * x31 + 0.0032 * x32 + 0.002725 * x33$$
$$+ 0.001183 * x41 + 0.0013 * x42 + 0.001103 * x43$$
$$+ 0.007813 * x51 + 0.007495 * x52 + 0.0076 * x53;$$
$$-y1 - y2 - y3 = R;$$
$$x11 + x12 + x13 + x21 + x22 + x23 + x31 + x32 + x33 + x41 + x42 + x43 + x51$$
$$+ x52 + x53 <= 15;$$
$$X11 + x21 + x31 + x41 + x51 - e1 <= 3;$$
$$x12 + x22 + x32 + x42 + x52 - e2 <= 3;$$

```
X13 + x23 + X33 + X43 + x53 − ℯ3 <= 3;
X11 + x21 + x31 + x41 + x51 >= 2;
X12 + X22 + X32 + X42 + X52 >= 2;
X13 + x23 + x33 + x43 + x53 >= 2;
```

Where R is a "satisficing level for f2". Then, "by graphically [and numerically] minimizing and maximizing f₂ over X, the feasible region defined by the original constraint set, we are able to find all the N-points.

The formulation of the linear program limits decision variables, $X_{jk}$ and $Y_k$, to integer values. Specifically, $X_{jk}$ must be equal to zero or one, while $Y_k$ may take any positive integer value less than or equal to the number of receiving stations on the network.

For this toy problem, we utilized Lindo Systems' software, Lingo, to input and solve this linear problem. The values of R that we used ranged in value from 0 to -6. The detailed solution to this problem is presented in the solutions section and compared to the results from some of the other previous thesis papers.

## 3. Solution of Constraint Reduced Method

As the solution procedure explained, we were able to run the model and obtain similar results mentioned in the references. In the appendices part, the Lingo program for the constraint reduced method for the toy problem and different values of R, ranging from 0 to -6, is provided. The N-Points acquired from this solution are tabulated in Table 7 and Table 8 below.

**Table 7.** N-points

| R Values | X-space, N-points | | | | | | | | | | | |
|---|---|---|---|---|---|---|---|---|---|---|---|---|
| **R** | *X* | *X1* | *X1* | *X2* | *X2* | *X2* | *X3* | *X3* | *X3* | *X4* | *X4* | *X4* |
| -6 | 1 | 1 | 1 | 1 | 1 | 1 | 1 | 1 | 1 | 1 | 1 | 1 |
| -5 | 1 | 1 | 1 | 1 | 1 | 1 | 1 | 1 | 1 | 1 | 1 | 0 |
| -4 | 1 | 1 | 1 | 1 | 1 | 1 | 1 | 1 | 1 | 0 | 1 | 0 |
| -3 | 1 | 1 | 1 | 1 | 1 | 1 | 1 | 1 | 1 | 0 | 0 | 0 |
| -2 | 1 | 1 | 1 | 1 | 1 | 1 | 0 | 1 | 1 | 0 | 0 | 0 |
| -1 | 1 | 1 | 1 | 1 | 1 | 1 | 0 | 1 | 0 | 0 | 0 | 0 |
| 0 | 1 | 1 | 1 | 1 | 1 | 1 | 0 | 0 | 0 | 0 | 0 | 0 |

**Table 8.** N-points continued

| X-space, N-points Continued | | | | | | | | | Y-space, N-Points | |
|---|---|---|---|---|---|---|---|---|---|---|
| **X** | **X** | **X** | **Y** | **Y** | **Y** | **E** | **E** | **E3** | **f₁(x)** | **f₂(x)** |

| 1 | 1 | 1 | 0 | 0 | 0 | 2 | 2 | 2 | 0.069601 | -6 |
|---|---|---|---|---|---|---|---|---|----------|----|
| 1 | 1 | 1 | 0 | 0 | 0 | 2 | 2 | 1 | 0.068498 | -5 |
| 1 | 1 | 1 | 0 | 0 | 0 | 1 | 2 | 1 | 0.067315 | -4 |
| 1 | 1 | 1 | 0 | 0 | 0 | 1 | 1 | 1 | 0.066015 | -3 |
| 1 | 1 | 1 | 0 | 0 | 0 | 0 | 1 | 1 | 0.064077 | -2 |
| 1 | 1 | 1 | 0 | 0 | 0 | 0 | 1 | 0 | 0.061352 | -1 |
| 1 | 1 | 1 | 0 | 0 | 0 | 0 | 0 | 0 | 0.058152 | 0 |

The following solution is gained through the thesis's results, and it is presented that the Lingo's output is closely matched the EVAL computer software developed by DOD.

```
        X space N-points                  Y space N-points

r₂      (x₁₁, ..., x₅₃, e₁, e₂, e₃)        (f₁(X), f₂(X))
------------------------------------------------------------
-6      (1,1,1,1,1,1,1,1,1,1,1,1,1,1,1,2,2,2)   (0.069602, -6) A

-5      (1,1,1,1,1,1,1,1,1,1,1,0,1,1,1,2,2,1)   (0.068499, -5) B

-4      (1,1,1,1,1,1,1,1,1,0,1,0,1,1,1,1,2,1)   (0.067316, -4) C

-3      (1,1,1,1,1,1,1,1,1,0,0,0,1,1,1,1,1,1)   (0.066016, -3) D

-2      (1,1,1,1,1,1,0,1,1,0,0,0,1,1,1,0,1,1)   (0.064078, -2) E

-1      (1,1,1,1,1,1,0,1,0,0,0,0,1,1,1,0,1,0)   (0.061353, -1) F

 0      (1,1,1,1,1,1,0,0,0,0,0,0,1,1,1,0,0,0)   (0.058153,  0) G
```

**Figure 7.** Solution

The following graphical representation depicts the Y-space which is the optimal values given from the trade-offs between two objective functions.

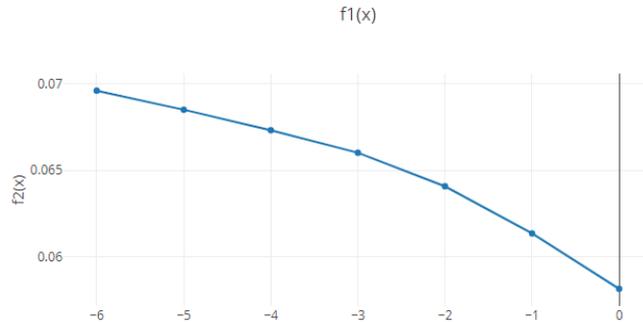

**Figure 2.** Graph of N-points in Y-space demonstrating the efficient frontier

## 4. Conclusion

This paper used a priori optimization to demonstrate the optimal assignment of HFDF receivers to the Generalized Search and Rescue (GSAR) network, which is independent of the weighting of the transmitter areas. The model objective presented was to optimize the estimated number of LOBs for HFDF receivers and to provide a reasonable share of the number of HFDF receivers allowed to cover the frequency. Although optimization models are of remarkable importance when it boils down to accuracy, being time consuming and engaging computational resources are the reasons to consider artificial intelligence approaches too, such as Simulated Annealing algorithm (Arani et al. 2021), Genetic Algorithm (Abir et al. 2020).